# Compact ultra-broadband light coupling on chip via nonadiabatic pumping

Weiwei Liu[1,#], Chijun Li[1,#], Bing Wang[1,*], Tianyan Chai[1], Lingzhi Zheng[1], Zhuoxiong Liu[1], Haoru Zhang[1], Shuaifei Ren[1], Xiaohong Li[2], Cheng Zeng[1,*], Jinsong Xia[1], Peixiang Lu[1,2,*]

[1]Wuhan National Laboratory for Optoelectronics and School of Physics, Huazhong University of Science and Technology, Wuhan 430074, China

[2]Hubei Key Laboratory of Optical Information and Pattern Recognition, Wuhan Institute of Technology, Wuhan 430205, China

[#]These authors contributed equally to this work

[*]Corresponding authors: wangbing@hust.edu.cn, zengchengwuli@hust.edu.cn, lupeixiang@hust.edu.cn

**Abstract**

Enlarging bandwidth capacity of the integrated photonic systems demands efficient and broadband light coupling among optical elements, which has been a vital issue in integrated photonics. Here, we have developed a compact ultra-broadband light coupling strategy based on nonadiabatic pumping in coupled optical waveguides, and experimentally demonstrated the designs in thin-film lithium niobate on insulator (LNOI) platform. We found that nonadiabatic transition would produce a decreased dispersion of the phases related to eigenstates in the waveguides. As a consequence, we realized high-efficiency directional transfer between edgestates for various wavelengths covering a 1-dB bandwidth of ~320 nm in experiment (>400 nm in simulation), with a coupling length (~50 μm) approximately 1/10 of that required in the adiabatic regime. Furthermore, we have constructed complex functional devices including beamsplitter and multiple-level cascaded networks for broadband light routing and splitting. Our work preserves significant advantages simultaneously in extending the operation bandwidth and minimizing the footprint, which demonstrates great potential for large-scale and compact photonic integration on chip.

**Introduction**

The rapid advancement of integrated photonics has opened the way for massive and high-speed data transfer on chip. These unique advantages lead to extensive applications including optical communications (*1, 2*), optical computing (*3-5*), quantum information processing (*6, 7*), and so on. To realize compact photonic integrated circuits, light coupling among waveguides plays an important role in light transmission and optical interconnection. However, due to serious material and waveguide dispersion, light coupling is highly sensitive to structure and wavelength. As a consequence, the response bandwidth and data capacity will be severely restricted by the existing bottlenecks. The urgent need to enlarge bandwidth capacity of the integrated photonic systems demands efficient and broadband light coupling among optical elements. Currently, a variety of effective approaches have been proposed and demonstrated for achieving the desired light coupling, including adiabatic transfer (*8, 9*), phase matching (*10*), dispersion control (*11*) and topological protection (*12*). One notable example is the adiabatic passage transfer (*8, 13*), which allows selective population transfer between coupling waveguides via an intermediate state, and is insensitive to the coupling strength in the adiabatic limit. However, it requires a long coupling configuration to ensure efficient adiabatic transfer, thus resulting in highly elongated waveguides and decreased integration density (*9*). More recently, artificial gauge field are developed to engineer the coupling dispersion among waveguides (*14*), which are beneficial for realizing broadband dispersionless coupling to certain extents. Nevertheless, most designs still suffer from challenges of covering a limited optical communication bands. Therefore, the realization of high-performance light coupling with ultra-broad bandwidth, high coupling efficiency and small interaction length still remains challenging.

As one of the most intriguing effects in quantum mechanics, topological pumping describes quantized transport of charge through an adiabatic cyclic evolution of the underlying Hamiltonian (*15*). Because of the worthwhile topological features, it has garnered significant attention from diverse platforms such as semiconductor quantum dots (*16*), cold atoms (*17, 18*), quasicrystals (*19, 20*), acoustics (*21*), quantum gas (*22*) and synthetic-dimension systems (*23, 24*). In particularly, coupled optical waveguides have offered an ideal photonic platforms for exploring topological pumping by constructing a cycle in the parameters space (*25-28*). Adiabatic transport of edge states in waveguide arrays has been demonstrated utilizing this interesting mechanism, and it provides a favorable approach for realizing broadband light coupling with robustness which is immune to structural perturbations and fabrication disorders (*29*). Moreover, it has been demonstrated that the topologically protected transfer can still happen with half-cycle topological pumping in finite lattices (*30*), and fast Thouless pumps was realized by non-Hermitian Floquet engineering in which non-adiabatic transitions were suppressed with time-periodic dissipation(*31*). Nevertheless, edge state transfers based on adiabatic evolution and Floquet-driven system both request a large interaction length, which imposes a trade-off between the response bandwidth and small footprint. The nonadiabatic effects (*21, 32, 33*), exhibiting tremendous distinctions in the interaction mechanism and showing great potential for resolving the contradictions, requires further exploration.

In this work, a compact and ultra-broadband light coupling strategy is developed based on nonadiabatic pumping in optical waveguide arrays, and the designs are further experimentally demonstrated in thin-film lithium niobate on insulator (LNOI) platform. Nonadiabatic pumping is created in finite off-diagonal Aubre-Andre-Harper (AAH) modeled photonic waveguides away from the adiabatic limitation. In this regime, a high-efficiency directional transfer of the incident light to the other edge can happen for various wavelengths covering a 1-dB bandwidth of ~320 nm in experiment (>400 nm in simulation), and the light coupling length is approximately 1/10 of that required for adiabatic transfer. Furthermore, broadband beamsplitter and multiple-level cascaded networks are constructed for light routings. This result preserves significant advantages simultaneously in extending the operation bandwidth and minimizing the footprint, which demonstrates great potential for large-scale and compact photonic integration on chip.

**Results**

We start by considering the one-dimensional (1D) off-diagonal Aubre-Andre-Harper (AAH) model, which can be also connected with the topological pumping of a Su-Schrieffer-Heeger (SSH) chain with $N$ of sites [$N$ is odd, Fig. 1A] (*34*). The system composes of a series of single mode waveguides that represent sites of a tight binding lattice. Topological pumping can be realized by varying the coupling strengths $J_1$ and $J_2$ between neighboring lattice sites. Light propagation in the waveguide array can be described by the tight-binding Hamiltonian

$$H = t\sum_{n=1}^{N} c_n^\dagger c_n + \sum_{n=1}^{N}[J_0 + (-1)^n \Delta J(z)]c_n^\dagger c_{n+1} + h.c. \qquad (1)$$

Here $c_n^\dagger$ and $c_n$ are the creator and annihilator on lattice site $n$ respectively. $t$ represents the on-site potential, which is a constant for the off-diagonal AAH model. The second term in Eq. (1) represents the couplings between nearest-neighbor waveguides with a constant (staggered) coupling strengths $J_0$ [$\Delta J(z)$], which could be manipulated by adjusting the waveguide separations. The distance between neighboring waveguides are defined as $d = d_0 + \Delta d\cos(2\pi z/P)$ [$d = d_0 - \Delta d\cos(2\pi z/P)$], where $z$ is the propagation distance and $P = 2L$ is the modulation period. Correspondingly, the coupling strengths can be approximately expressed as $J_1 = J_0 - \Delta J \cos(2\pi z/P)$ [$J_2 = J_0 + \Delta J \cos(2\pi z/P)$] (See Supplemental Material). Since the edge states transfer can happen with half-cycle pumping, the total length of the waveguide array is set to $L$, as illustrated in the bottom panel of Fig. 1A. By diagonalizing the Hamiltonian in Eq. (1), we get the energy spectrum of this system, as shown in Fig. 1B. There is exactly a zero-energy eigenstate lying in the band gap, which is an edge state with light localized at the left edge for $z = 0$, as clearly displayed in the top panel of Fig. 1B. With the increases of propagation distances $z$, the band gap reduces gradually and nearly closes at $z = L/2$ ($\Delta J = 0$). Correspondingly, the localized state would be diverged at this gap-closing point, and then fully transferred to the right edge of the waveguide array for $z = L$. Moreover, this feature of the state transfer is essentially independent on the number of the lattice sites (waveguides), as the edge state still remains for the least lattice of $N = 3$. One critical issue for the adiabatic pumping of the edge states lies on that the energy

transfer should be fully adiabatic all over this process. The adiabatic condition can be expressed as (30)

$$\eta = \frac{\sqrt{\Delta J \Omega}}{\Delta E} \ll 1, \quad (2)$$

where $\Delta J$ is the variation amplitude of the coupling strength, and $\Omega$ is the spatial modulation frequency, $\Omega = 2\pi/P$. $\Delta E$ represents the energy gap between the adjacent states. The adiabatic condition is quantitatively analyzed by adopting a waveguide array with $N = 3$ for instance. Thin-film LNOI is selected to construct the coupled waveguide structure, due to its wide transparency window and small material dispersion (35). Fig. 1C maps $\eta$ as function of the waveguide length and wavelength. It indicates that an interaction length larger than ~200 μm is required to realize adiabatic transfer over a broad band ($\eta = 0.3$), and a much longer length over ~500 μm is required to achieve a high transfer efficiency (See Supplemental Material), which is not beneficial for achieving compact light coupling on chip.

In order to obtain a broadband and efficient excitation transfer with compact waveguides, light propagation in the waveguides is investigated under a nonadiabatic regime. With the paraxial approximation, light transfer in coupled waveguides can be calculated by the following coupled mode equation

$$i\frac{\partial \Psi}{\partial z} = H\Psi, \quad (3)$$

Where $H$ is the systematic Hamiltonian descript by Eq. (1) for $N = 3$. $\Psi$ is the complex amplitude of the electric field in the waveguides. Assumed that an edgestate of $\Psi(0) = [1, 0, 0]^T$ is incident from input port of the waveguides. The incident field can be decomposed as $\Psi(0) = a_1(0)\Psi_1(0) + a_2(0)\Psi_2(0) + a_3(0)\Psi_3(0)$, where $\Psi_n(0)$ ($n = 1, 2, 3$) represent the eigenstates of the coupled waveguides at $z = 0$. In the adiabatic process, transfer of the states happens with corresponding weight coefficients of the eigenstates remaining unchanged ($[a_1(0), a_2(0), a_3(0)]$). However, in the nonadiabatic regime, the eigenstates corresponding to different bands would transit to each other, resulting in propagation-dependent weight coefficients in the evolution (34), i.e., $[a_1(z), a_2(z), a_3(z)]$. Correspondingly, the field distribution in the waveguides can be expressed as

$$\Psi(z) = \left[ a_1(z)\Psi_1(z)e^{-i\Delta\varphi} + a_2(z)\Psi_2(z) + a_3(z)\Psi_3(z)e^{i\Delta\varphi} \right]e^{i\varphi}, \quad (4)$$

where $\varphi = \int\beta(z)dz$ representing the propagation phase of the zero-energy eigenstate. $\Delta\varphi$ is the phase contrast between the adjacent modes, which originates from the mismatch of propagation constants and nonadiabatic transition (See Supplemental Material).

The coupled mode equation Eq. (3) are exactly solved by numerical method to explore the light transfer process. Fig. 2A shows the normalized field intensity (transmission) in the corresponding edge waveguide of output port, monitoring as a function of the waveguide length $L$. In the adiabatic region, the transmission exhibits strong oscillation versus $L$, and the numerically solved results are consistent with the analytic solution, which can be attributed to the phase mismatching and interference of the eigenstates (See Supplemental Material). In contrast, away from the adiabatic region, the transmission increases sharply first with the waveguide length, and we can still find a high transmission peak at around $L = 50$ μm, indicating a high-efficiency transfer of the incident edgestate. Eq. (4) indicates that the accumulated phases contrast play an important role for the superposition of eigenstates in the coupled waveguides. Fig. 2B (red curve) presents the accumulated phases contrast ($\Delta\varphi$) at $z = L$. It is close to $\Delta\varphi = 2\pi$, thus leading to a constructive superposition of the eigenstates. More interestingly, Fig. 2B also displays that the accumulated phase contrast during nonadiabatic transfer varies slightly with the wavelength, implying a decreased dispersion of the nonadiabatic pumping process. For comparison, we also calculated the phase contrast between the even and odd eigenstates of two coupled waveguide with an identical length, which represents the typical structure of a conventional directional coupler (DC). In this structure, the accumulated phase contrast is proportional to the coupling strength, $\Delta\varphi = (\beta_{even} - \beta_{odd})L = 2J_0L$, where $J_0$ represents the constant coupling strength between the two waveguide (*29*). Fig. 2B (black curve) indicates that $\Delta\varphi$ varies strongly with wavelength, implying a serious dispersion of the conventional DC structure. Importantly, the decreased dispersion due to nonadiabatic transition is advantageous for realizing a high-efficiency and broadband transfer of the edgestate. To confirm this prediction, Fig. 2C displays a continuous mapping of the transmission as functions of the waveguide length $L$ and wavelength. It can be notably observed that the transmission is nearly unity and experiences little dispersion around $L = 50$ μm, all over the calculated wavelength range from 1.3 μm to 1.7 μm. Therefore, it demonstrates an effective approach for achieving ultra-broadband light transfer of the edgestate, with an interaction length less than 1/10 of that required in the adiabatic regime (>500 µm, See Supplemental Material).

To verify the theoretical results in practice, waveguide structures are designed and fabricated based on X-cut thin-film LNOI (300-nm LN layer). LN ridge waveguides are designed to have a top width of 650 nm, and the etch depth is 200 nm with a remaining slab of 100 nm, which allows for fundamental transverse electric like (TE) mode around 1550 nm. The separation between the top and middle waveguides is set as $d = d_0 + \Delta d \cos(2\pi z/P)$, with $d_0 = 420$ nm, $\Delta d = 200$ nm. Silicon dioxide film is covered on the device for cladding. Three-dimensional (3D) full-wave simulations are performed using a commercial finite-difference time-domain analysis solver (Lumerical FDTD). Fig. 2C presents simulated field distributions ($|E|^2$) in the LN waveguides for the nonadiabatic pumping structure. The whole length of the interaction region ($L$) is optimized to be ~50 µm. One can observe that the input state can be highly transferred to opposite edge at the output port, and simulated results are consistent with the theoretical prediction that light distribution in the intermediate waveguide suggest a nonadiabatic transfer process. More importantly, the high-efficiency transfer of edgestate happens in a broadband spanning from 1.3 µm to 1.7 µm, suggesting an operation bandwidth over 400 nm. In contrast, for conventional two-waveguide directional coupler (DC) with an identical interaction length, light coupling only exhibits a narrowband response from 1.5 µm to 1.6 µm, as shown in Fig. 2D.

For experiment, LN waveguides were fabricated using standard nanofabrication process (See Supplemental Material). Fig. 3A shows the microscopic image and corresponding enlarged parts of the fabricated samples, with a conventional DC structure for comparison. Near-infrared lasers covering the optical communication bands (1330-1650 nm) are used as light sources. The input and output ports are both tapered and polished, and lens fibers are used for efficiently coupling single-frequency laser with various wavelengths into (out of) the chip. A high-performance spectrometer (AQ6370B, Yokogawa) is used to measure the transmission spectra, and a near-infrared camera (991SWIR, Artray) is used to image the light spots from end facet of the waveguides under a microscope. Fig. 3B shows experimentally measured output spots of the nonadiabatic pumping device and conventional DC respectively, at wavelengths of 1330, 1390, 1450,

1550, and 1650 nm. As observed, for the nonadiabatic device, the input signal ($I_2$) can be almost transferred to the crossing output port ($O_1$), with nearly none optical signal outgoing from the other port ($O_2$). For comparison, the coupling property of the conventional DC varies strongly at different wavelengths. Fig. 3C presents the coupling ratio from the input to the output port (referred to an individual waveguide) of the devices. The nonadiabatic-pumping structure possess a 1-dB bandwidth in the whole measured range from 1330 nm to 1650 nm, indicating an practical operation bandwidth of at least ~320 nm (>400 nm in simulation), which is ~2 times of the conventional DC (~160 nm). In addition, Fig. 3D shows that the device exhibits a general insertion loss (compared to an individual LN waveguide) less than 0.5 dB and a low crosstalk less than -10 dB, indicating a high performance of the designed light coupling structure via nonadiabatic pumping.

Furthermore, we also designed and fabricated various nonadiabatic-pumping based structure with different waveguide widths and separations. Numerical and experimental results show that ~100 nm variation of waveguide width and separation can have slight influence on the performance of the ultra-broadband DC (See Supplemental Material). Though the samples are fabricated based on the thin-film LN platform, we would like to stress that the basic principle is independent on materials, which can be verified by the numerically simulated results for silicon waveguides (See Supplemental Material). Table S1 summarizes the performances of typical broadband coupling designs on chip. The operation bandwidth of our nonadiabatic pumping design is 2-3 times broader than the other designs with similar interaction lengths (*12, 14, 29*). It should be noted that, although adiabatic passage transfer enables to provide a very broad coupling bandwidth, it requests a much longer interaction length on the order of millimeters, which is about 20 times larger than the nonadiabatic design (*9*). Therefore, our design preserves significant advantages simultaneously in extending the operation bandwidth and minimizing the footprint. The ultra-broadband operation bandwidth, great robustness and compatibility of this nonadiabatic pumping design shows great potential for efficient light coupling and interconnection in versatile compact platforms.

In order to demonstrate the promise of the ultra-broadband light coupling in integrated photonics, the nonadiabatic pumping structures are constructed to form complex functional devices for light routing. Fig. 4A presents a microscopic image of a 3-dB beamsplitter produced by symmetrically aligning the nonadiabatic-pumping based coupling elements. The enlarged part on right panel of Fig. 4A shows a uniform distribution of the two sides. As a consequence, the input signal can be equally split into two routes at different wavelengths, as shown in Fig. 4B. Correspondingly, Fig. 4C presents quantitatively simulated and measured coupling ratio and splitting ratio of the beamsplitter. The coupling ratio distributes in a range of (-3±0.5) dB, and the splitting ratio of the two ports ($I_{O1}/I_{O2}$) remains nearly to be 1:1, for the whole measured wavelengths from 1330 nm to 1650 nm. It means that the broadband performance can still maintain for combined functional elements. Furthermore, the splitter elements are connected to form multiple-level cascaded networks, including two-level (4-port, Figs. 4D-4E) and three-level (8-port, Fig. 4F) configurations. In the cascaded networks, the input light can be equally routed to all of the output ports with a same broadband response. It demonstrates that our broadband coupling strategy via nonadiabatic pumping is not only effective for individual element, but also powerful for constructing cascaded networks, showing great prospects for large-scale and compact photonic integration on chip.

**Discussion**

A compact and ultra-broadband light coupling strategy has been developed via nonadiabatic pumping in coupled optical waveguides. The proposal has been successfully demonstrated in practice in a typical thin-film LNOI platform, which enables to realize ultra-broadband and efficient light coupling spanning from 1330 nm to 1650 nm (1-dB operation bandwidth of ~320 nm), as well as with a small interaction length (50 μm) approximately 1/10 of the adiabatic transfer regime. For functional applications, this light coupling strategy is of remarkable convenience for constructing broadband beamsplitter and multiple-level cascaded networks for compact light routings on chip. More importantly, it has been demonstrated that the basic principle is independent on materials and can be realized on versatile compact platforms. The nonadiabatic pumping designs preserve significant advantages simultaneously in extending the operation bandwidth and minimizing the footprint, which shows great potential for large-scale and compact photonic integration on chip.

**Materials and Methods**

**Device design and simulation.** LN waveguide devices are designed based on X-cut thin-film LN on insulator (300-nm LN layer). The ridge waveguides has an etch depth of 200 nm, with a sidewall angle of ~70° according to previous results(*36*), which allows for fundamental transverse-electric (TE) like mode around 1550 nm. The waveguides are covered with silicon dioxide film for cladding. The modal effective index and light propagation in the waveguide devices are simulated using three-dimensional (3D) full-wave simulations with a commercial finite-difference time-domain analysis solver (Lumerical FDTD). The refractive indices of $LiNbO_3$ and Si are referred to reported results (*37, 38*).

**Device fabrication.** The designed samples are fabricated on a commercial X-cut LNOI wafer on silicon substrate (NanoLN). The LNOI wafer was patterned using electron beam lithography (EBL), and then the structures ware etched using Ar plasma in an inductively coupled plasma (ICP) reactive ion etching machine. Subsequently, a layer of $SiO_2$ with a thickness of 1000 nm was deposited using plasma enhanced chemical vapor deposition (PECVD). Finally, grinding and polishing processes were performed to expose the coupling facets for lens fiber coupling.

**Experimental measurements.** For broadband optical measurements, a tunable continuous-wave (CW) telecom-band laser (1500-1630 nm, TSL-510, Santec) and near-infrared distributed feedback (DFB) lasers covering other individual wavelengths (1330, 1370, 1390, 1410, 1430, 1450, 1470, and 1650 nm) are used as light sources. The polarization of incident light was controlled by a fiber polarization controller (FPC032, Thorlabs). The signal was efficiently coupled into (out of) the chip using lens fibers. A commercial 10:90 Y-splitter was connected to split the output signal into two paths. One terminated (10%) to an optical power meter for monitoring and tuning the alignment between lens fibers and output ports. The other one (90%) was connected to a high-performance optical spectrum analyzer (AQ6370B, Yokogawa) for transmission spectra measurement. The output spots were monitored and imaged using a near-infrared camera (991SWIR, Artray) under a microscope. Besides, several reference circuits including only individual waveguides with tapered ports were fabricated on the same chip to calibrate the transmission efficiency. The experimental setup for optical measurements can be seen in Supplementary Note 2.

**Acknowledgments**

**Funding:** This research is supported by the National Natural Science Foundation of China (No. 62375097, No. 12374305, No. 12204363, No. 12021004); the Natural Science Foundation of Hubei Province (No. 2023AFB822, No. 2020CFA004); Hubei Key Laboratory of Optical Information and Pattern Recognition, Wuhan Institute of Technology (No. 202202). The authors gratefully acknowledged the Center of Micro-Fabrication and Characterization (CMFC) of WNLO (HUST) for their support in the nanofabrication (EBL and ICP) of devices.

**Author contributions:** W. L., B. W. and P. L conceived the idea and guided the project. W. L., C. L, T. C, C. Z. and B. W. designed and fabricated the devices. W. L. and T. C performed the theoretical analysis, numerical simulations and experimental measurements. All the authors contributed to discussion and writing of the manuscript.

**Competing interests:** The authors declare no competing interests.

**Data and materials availability:** All data needed to evaluate the conclusions in the paper are present in the paper and/or the Supplementary Materials.

# Figures

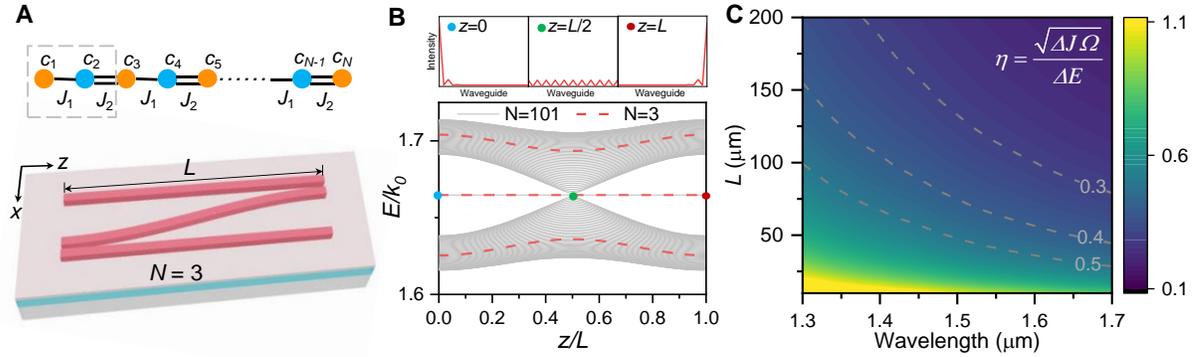

**Fig. 1. Topological pumping in optical waveguides.** (**A**) Schematic of a SSH chain with an odd number of sites and a waveguide array ($N = 3$) for topological pumping of edge states. (**B**) Energy spectra versus the normalized propagation length ($z/L$) in a chain comprising $N = 101$ and $N = 3$ sites, respectively. Up panel shows eigenstates of the chain at $z = 0$, $L/2$ and $L$. (**C**) Mapping of the adiabatic condition factor as functions of wavelength and waveguide length.

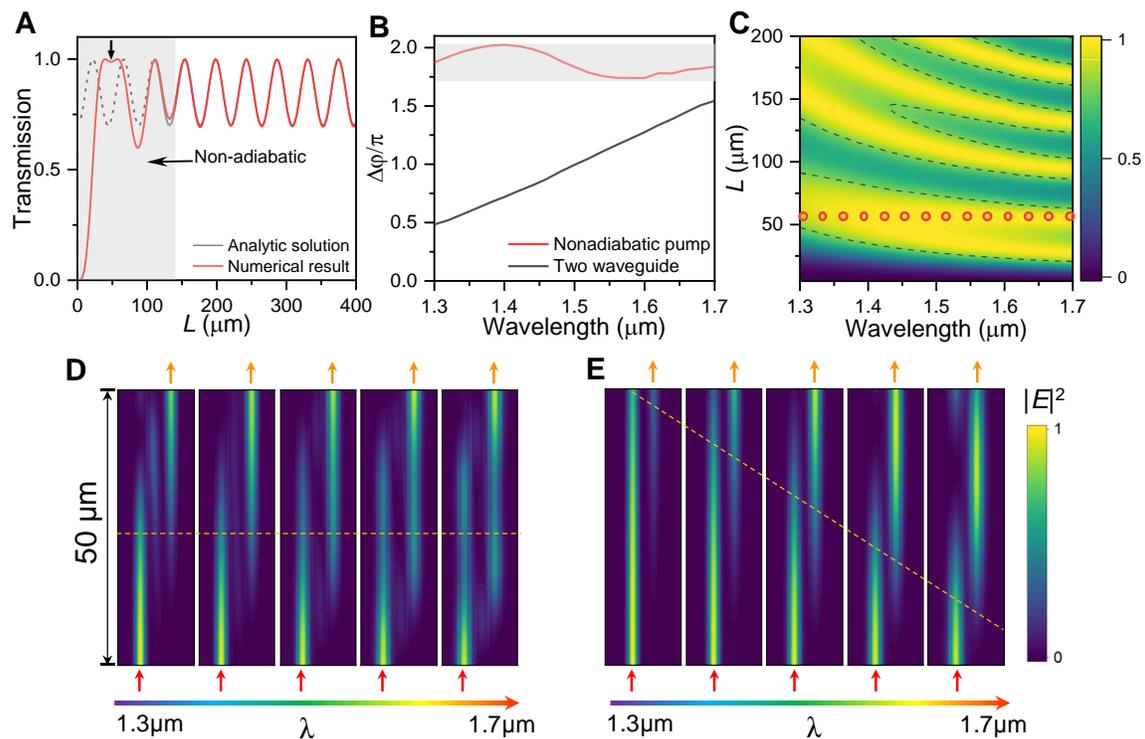

**Fig. 2. Nonadiabatic pumping and transfer of edgestate.** (**A**) Normalized field intensities (transmissions) at the output port as a function of the waveguide length, solved by numerical method and analytic solution respectively. (**B**) Plots of the phase contrast $\Delta\varphi$ between adjacent eigenstates at the output port, by considering the total accumulated phase (red curve) of the nonadiabatic pumping transition, and that of the two waveguide structure (black curve). (**C**) Mapping of transmission as functions of waveguide length and wavelength. The contour lines represent a transmission larger than 0.8 (-1 dB). (**D** and **E**) Simulated field distributions ($|E|^2$) in the LN waveguides of the nonadiabatic pumping design and conventional DC composed of two coupled waveguides, respectively.

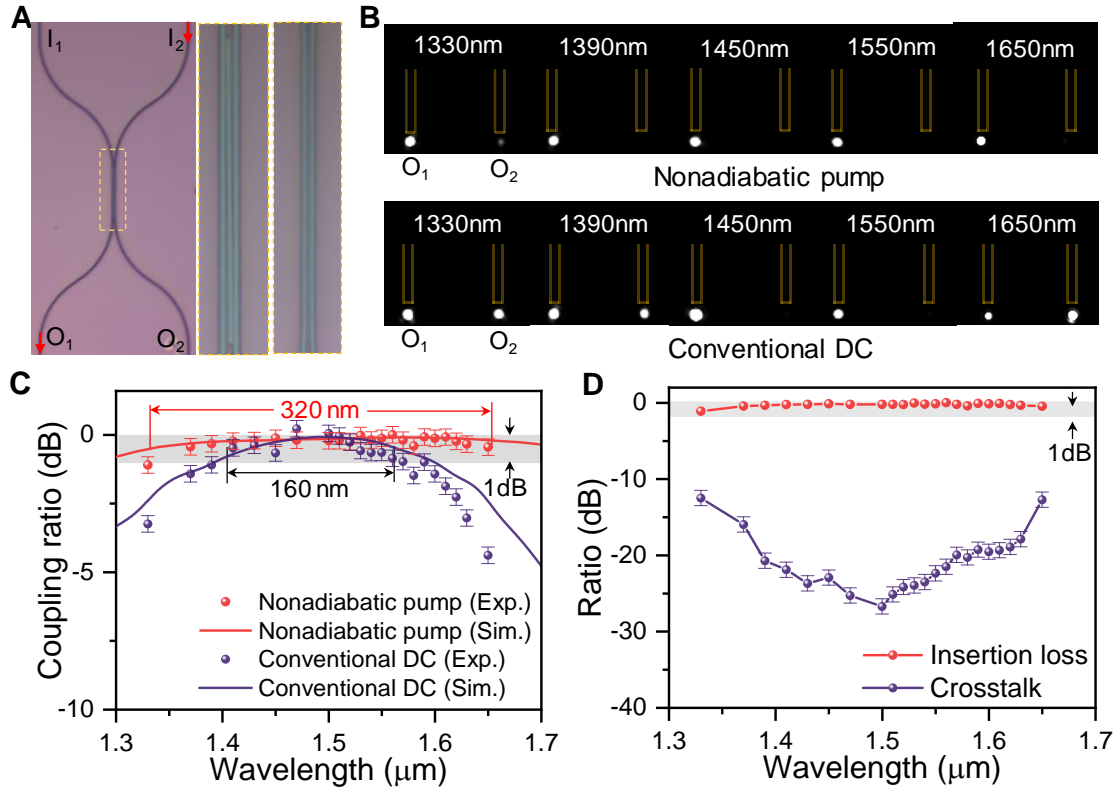

**Fig. 3. Experimental demonstration of ultra-broadband light coupling on chip.** (**A**) Microscopic image of the samples on a LNOI platform. (**B**) Experimentally imaged scattering fields from output ports of the nonadiabatic pumping device and conventional DC respectively, at different wavelengths (1330, 1390, 1450, 1550, and 1650 nm). (**C**) Plots of measured and simulated coupling ratios for nonadiabatic pumping device and conventional DC respectively. (**D**) Plots of insertion loss and crosstalk of the nonadiabatic pumping device.

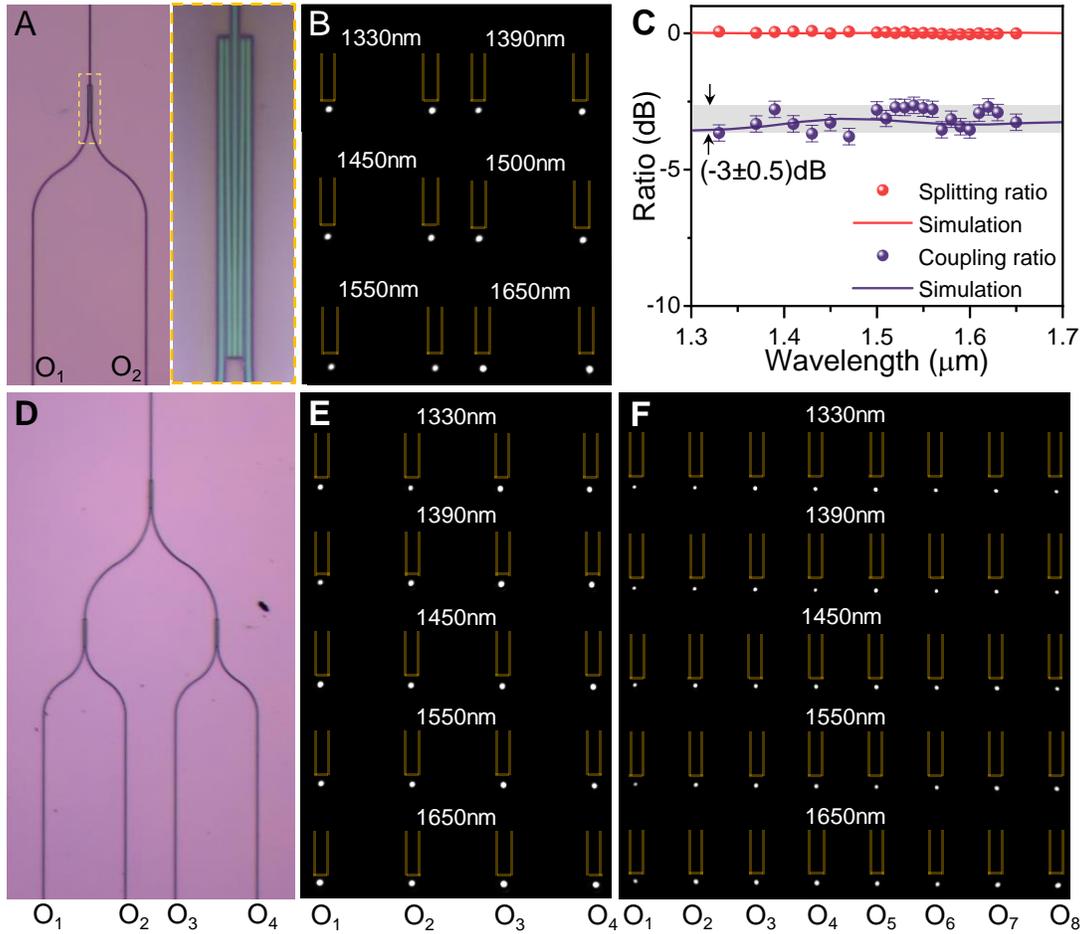

**Fig. 4. Broadband light routings based on nonadiabatic pumping design.** (**A**) Microscopic image of a 3-dB beamsplitter fabricated based on the nonadiabatic pumping coupler. (**B**) Experimentally imaged scattering fields from output ports of the beamsplitter at different wavelengths (1330, 1390, 1450, 1500, 1550, and 1650 nm). (**C**) Plots of measured and simulated coupling ratios and splitting ratios for the beamsplitter. (**D**) Microscopic image of a two-level (4-port) cascaded networks. (**E** and **F**) Experimentally imaged scattering fields from output ports of the two-level (4-port) and three-level (8-port) cascaded networks at different wavelengths (1330, 1390, 1450, 1550, and 1650 nm), respectively.